%% file: main.tex
\documentclass[]{pasj01}
\draft

\begin{document} 
\Received{}
\Accepted{}

\def\kms{km s$^{-1}$}
\def\vlsr{$v_{\rm LSR}$}

\def\Herschel{{\it Herschel}}
\def\Spitzer{{\it Spitzer}}
\def\WISE{{\it WISE}}
\def\AKARI{{\it AKARI}}

\def\COal{\atom{C}{}{12}\atom{O}{}{} ($J$=1--0)}
\def\CObl{\atom{C}{}{13}\atom{O}{}{} ($J$=1--0)}
\def\COcl{\atom{C}{}{}\atom{O}{}{18} ($J$=1--0)}

\def\COam{\atom{C}{}{12}\atom{O}{}{} ($J$=2--1)}
\def\CObm{\atom{C}{}{13}\atom{O}{}{} ($J$=2--1)}
\def\COcm{\atom{C}{}{}\atom{O}{}{18} ($J$=2--1)}

\def\COa{\atom{C}{}{12}\atom{O}{}{}}
\def\COb{\atom{C}{}{13}\atom{O}{}{}}
\def\COc{\atom{C}{}{}\atom{O}{}{18}}

\def\HII{H \emissiontype{II}}

\input{s0_title}
\maketitle
\input{s0_abstract}
\input{s1_introduction}

\input{s2_data}
\input{s3_results}
\input{s4_discussion}

\input{s5_conclusions}

\input{s9_references}

\end{document}

%% file: s0_title.tex
\title{FOREST Unbiased Galactic plane Imaging survey with the Nobeyama 45-m telescope (FUGIN) 2: Possible evidence for formation of NGC~6618 cluster in M17 by cloud-cloud collision}

\author{Atsushi \textsc{nishimura}\altaffilmark{1*}}
\author{Tetsuhiro \textsc{minamidani}\altaffilmark{2,3}}
\author{Tomofumi \textsc{umemoto}\altaffilmark{2,3}}
\author{Shinji \textsc{fujita}\altaffilmark{1}}
\author{Mitsuhiro \textsc{matsuo}\altaffilmark{2}}
\author{Yusuke \textsc{hattori}\altaffilmark{1}}
\author{Mikito \textsc{kohno}\altaffilmark{1}}
\author{Mitsuyoshi \textsc{yamagishi}\altaffilmark{4}}
\author{Yuya \textsc{tsuda}\altaffilmark{5}}
\author{Mika \textsc{kuriki}\altaffilmark{6}}
\author{Nario \textsc{kuno}\altaffilmark{6,7}}
\author{Kazufumi \textsc{torii}\altaffilmark{2,3}}
\author{Daichi \textsc{tsutsumi}\altaffilmark{1}}
\author{Kazuki \textsc{okawa}\altaffilmark{1}}
\author{Hidetoshi \textsc{sano}\altaffilmark{1,8}}
\author{Kengo \textsc{tachihara}\altaffilmark{1}}
\author{Akio \textsc{ohama}\altaffilmark{1}}
\author{Yasuo \textsc{fukui}\altaffilmark{1,8}}

\altaffiltext{1}{Department of Physics, Nagoya University, Furo-cho, Chikusa-ku, Nagoya, Aichi 464-8602, Japan}
\altaffiltext{2}{Nobeyama Radio Observatory, National Astronomical Observatory of Japan (NAOJ), National Institutes of Natural Sciences (NINS), 462-2 Nobeyama, Minamimaki, Minamisaku, Nagano 384-1305, Japan}
\altaffiltext{3}{Department of Astronomical Science, SOKENDAI (The Graduate University for Advanced Studies), 2-21-1 Osawa, Mitaka, Tokyo 181-8588, Japan}
\altaffiltext{4}{Institute of Space and Astronautical Science, Japan Aerospace Exploration Agency, Chuo-ku, Sagamihara 252-5210, Japan}
\altaffiltext{5}{Graduate School of Science and Engineering, Meisei University, 2-1-1 Hodokubo, Hino, Tokyo 191-0042, Japan}
\altaffiltext{6}{Department of Physics, University of Tsukuba, 1-1-1 Ten-nodai, Tsukuba, Ibaraki 305-8577, Japan}
\altaffiltext{7}{Center for Integrated Research in Fundamental Science and Technology (CiRfSE), University of Tsukuba, Tsukuba, Ibaraki 305-8571, Japan}
\altaffiltext{8}{Institute for Advanced Research, Nagoya University, Chikusa-ku, Nagoya, Aichi 464-8601, Japan}

\email{nishimura@a.phys.nagoya-u.ac.jp}


%% file: s0_abstract.tex
\begin{abstract} 

We present \COal, \CObl \ and \COcl \ images of the M17 giant molecular clouds obtained as part of FUGIN (FOREST Ultra-wide Galactic Plane Survey InNobeyama) project.
The observations cover the entire area of M17 SW and M17 N clouds at the highest angular resolution ($\sim$19$''$) to date which corresponds to $\sim$ 0.15 pc at the distance of 2.0 kpc.
We find that the region consists of four different velocity components: very low velocity (VLV) clump, low velocity component (LVC), main velocity component (MVC), and high velocity component (HVC).
The LVC and the HVC have cavities.
UV photons radiated from NGC 6618 cluster penetrate into the N cloud up to $\sim$ 5 pc through the cavities and interact with molecular gas.
This interaction is correlated with the distribution of YSOs in the N cloud.
The LVC and the HVC are distributed complementary after that the HVC is displaced by 0.8 pc toward the east-southeast direction, suggesting that collision of the LVC and the HVC create the cavities in both clouds.
The collision velocity and timescale are estimated to be 9.9 \kms \ and $1.1 \times 10^{5}$ yr, respectively.
The high collision velocity can provide the mass accretion rate up to 10$^{-3}$ $M_{\solar}$ yr$^{-1}$, and the high column density ($4 \times 10^{23}$ cm$^{-2}$) might result in massive cluster formation.
The scenario of cloud-cloud collision likely well explains the stellar population and its formation history of NGC 6618 cluster proposed by \citet{hof08}.

\end{abstract}

%% file: s1_introduction.tex
\section{Introduction}

High-mass stars are fundamental pieces in the Universe owing to its creation of heavy elements and release of energy into the interstellar medium.
However, its formation mechanism is still poorly understood, especially that early O-type stars ($> 32 M_{\solar}$) are discussed little to date (e.g., \cite{zin07}).
Theoretically, turbulent core accretion model \citep{mck03} and competitive accretion model \citep{bon01} are well investigated as the mass accretion processes of high-mass star formation (for a recent review, see \cite{tan14}).
However, these scenarios still lack detailed validation with the initial conditions provided by observations, since it was difficult to map the whole giant molecular cloud (GMC) at a sufficiently high angular resolution to resolve individual star forming regions due to the angular extent of nearby GMCs are large, more than a few several degrees (e.g., the Orion A cloud, \cite{nis15}).
The \Herschel \ space telescope successfully observed these regions in mid- to far-infrared dust continuum emission, although it was yet difficult to resolve the individual regions near the Galactic plane, where the sight-line contamination is significant \citep{mol16}.
Spectroscopic observations of molecular clouds have better capability to resolve the clouds by separating their velocities.

Recent spectroscopic observations of molecular clouds have discovered remnants of supersonic collisions of two clouds in super star cluster formation regions \citep{fur09, oha10, fuk14, fuk16}, high-mass stars formation regions (\cite{tor11, shi13, tor15, fuk15, tor17, fuk17a, fuk17b, fuk17c, hay17, nis17, oha17a, oha17b, san17, tsu17}) and low- and intermediate-mass stars formation regions \citep{nak14, gon17}.
Cloud-cloud collision was studied using numerical simulations \citep{hab92, ana10, ino13, tak14, haw15a, haw15b, wu15, bal15, wu17, bal17}.
These authors found that the cloud-cloud collision induces the formation of cloud cores by enhanced self-gravity as a consequence of shock compression.
\citet{ino13} shows in their magnetohydrodynamic (MHD) simulations on the cloud-cloud collision that colliding molecular gas can create dense and massive cloud cores in the shock-compressed interface and they suggest higher mass accretion rate ($\sim 10^{-3} M_{\solar}$ yr$^{-1}$), which should be necessary to form early O-type stars, can be achieved by enhanced effective sound speed.

In spite of recent vigorous studies, good candidates of cloud-cloud collision are still rare in observations.
The array heterodyne receivers like FOREST  \citep{min16}, the 4-beam SIS receiver equipped with the 45 m telescope at Nobeyama Radio Observatory, provide an ideal tool for the purpose.
Using FOREST receiver, we are carrying out the large scale Galactic plane survey FUGIN (FOREST  Unbiased Galactic plane Imaging survey with Nobeyama 45-m telescope; Umemoto et al. 2017).
A number of high-mass star formation regions and \HII \ regions are contained in the survey region ($10^{\circ} \leqq l \leqq 50^{\circ}$, $-1^{\circ} \leqq b \leqq 1^{\circ}$).
Since the highest angular resolution at $\sim 20''$ allows us for detailed study of high-mass star forming molecular clouds in the Galactic plane, FUGIN dataset accelerates the observational studies of cloud-cloud collision.

M17 is one of the best known sites of active high-mass star formation (e.g., \cite{chi08}) at a distance of 2.0 kpc \citep{chi16}. 
The region is heavily ionized by associated cluster NGC 6618 which contains $>53$ OB stars \citep{hof08} including O4-O4 binary system CEN1 \citep{chi80, rod12}.
In the M17 region the molecular clouds which are located around the \HII \ region including famous M17 SW cloud attracted the most intense attention in the previous molecular observations \citep{lad74, lad75, elm76, thr83, mar84, sne86, rai87, gue88, stu88, stu90, hob92, wan93, ber97, how00, wil03, per10, per12, per15a, per15b, yam16}, while only a few studies were conducted on the whole cloud including the northern part of the GMC \citep{wil99, pov09}.

We present large scale images of the M17 cloud with FOREST in the \COa, \COb \ and \COc \ $J=1-0$ transitions, and the results in this paper. 
Section 2 describes observational details and the data quality, and Section 3 gives the CO results including the distribution of the cloud and velocity and the derived physical parameters. 
Section 4 is devoted to discuss the star formation in M17 and Section 5 concludes the paper.

%% file: s2_data.tex
\section{Data}
\label{sec:data}
The data used in this paper is provided from Internal Release 1.3 (IR 1.3) of FUGIN project (Umemoto et al. submitted).
The data is taken between March 24 and May 27, 2014 by using the Nobeyama Radio Observatory 45 m telescope with a newly developed multi-beam receiver FOREST \citep{min16}.
The FX-type correlator SAM45 \citep{kun11, kam12} was used as the backend.
We observed \COal, \CObl, and \COcl \ transitions simultaneously.
The typical system temperature including an atmosphere noise is $T_{\rm sys} \sim$ 250 K for the 110 GHz band during the observation season. 
The pointing accuracy is better than 3 arcsec.
The on-the-fly (OTF) observations \citep{saw08} were carried out for each submap which is divided in 1 deg$^{2}$ size.
For intensity calibration, we observed M17 SW as a standard source to evaluate a scaling factor every day.

The observed data is processed by the NOSTAR reduction package to create a standard FITS image.
We fitted the baseline of each spectrum in the submaps by a polynomial function to subtract a baseline ripple.
After the baseline subtraction, we created the FITS image with an angular resolution of $\sim 20''$ and a velocity resolution of 1 km s$^{-1}$.
The typical noise levels $T_{\rm rms}$ are 0.9 K, 0.2 K and 0.2 K for frequency band of \COal \CObl \  and \COcl, respectively.
More detailed information about the dataset is described in Umemoto et al. (submitted).

%% file: s3_results.tex
\section{Results and analyses}
\label{sec:results}

\subsection{Spatial distributions}
\label{sec:cloud}

\begin{figure}[t]
 \begin{center}
  \includegraphics[width=0.7\textwidth]{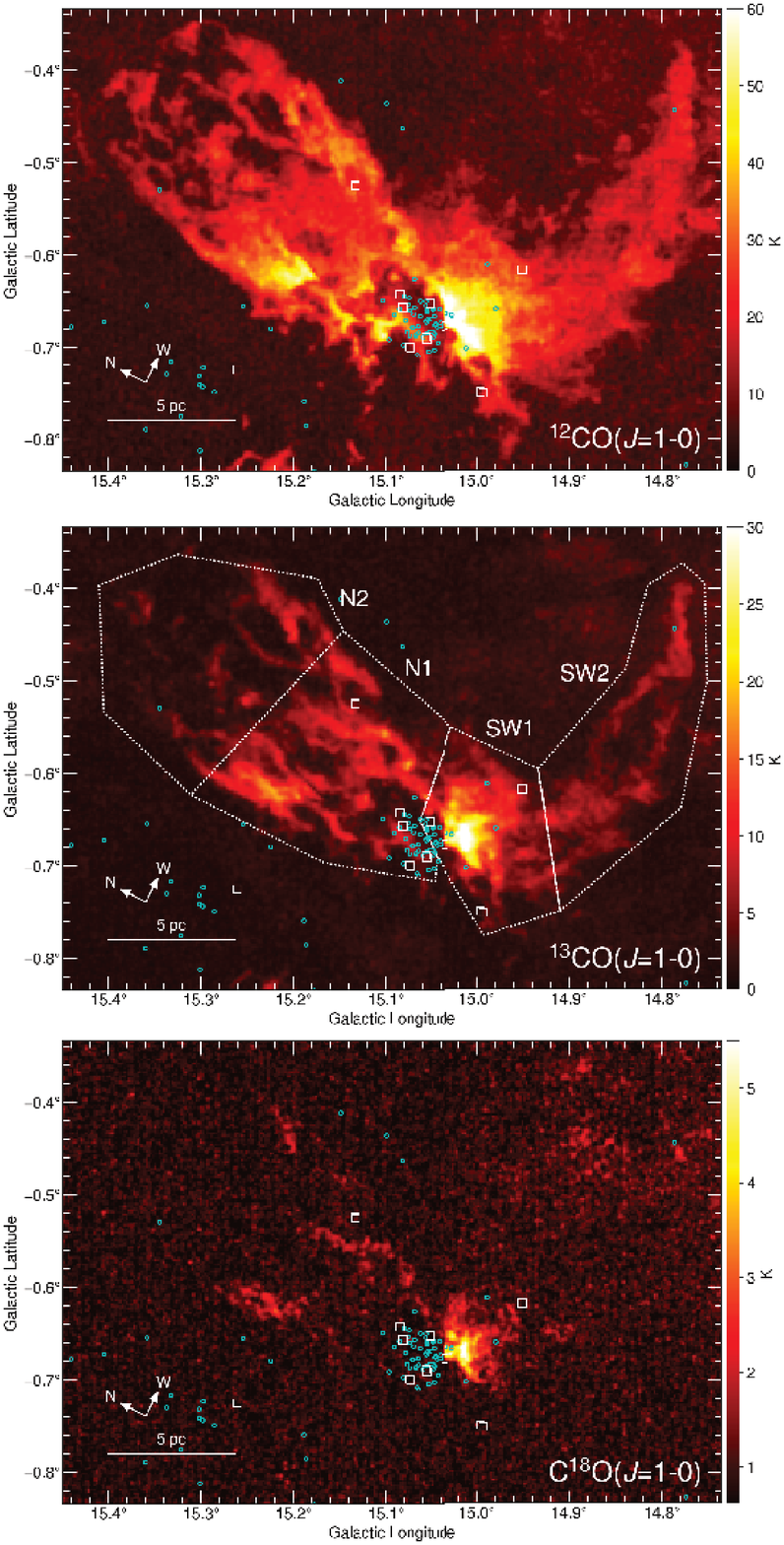} 
 \end{center}
\caption{Peak intensity maps of (a) \COal \ with peak intensity of 70.2 K, (b) \CObl \ with peak intensity of 35.0 K , and (c) \COcl \ with peak intensity of 5.9 K toward the M 17 molecular clouds. The velocity range used for the analyses is 0 km s$^{-1}$ $<$ \vlsr $<$ 40 km s$^{-1}$. Subregions SW1, SW2, N1 and N2 are indicated as dotted polygons. White squares and cyan circles show O-type and B-type stars \citep{hof08}.}
\label{fig:results_peak}
\end{figure}

Figure \ref{fig:results_peak} shows the peak intensity maps of \COal, \CObl \ and \COcl \ observed with Nobeyama 45 m telescope.
The M17 molecular cloud complex is fully covered with improved angular resolution at 19$''$ compared with a previous observation of \COam \ and \CObm \ at 32$''$ resolution \citep{pov09}.
As many authors previously noted, the region mainly consists of two clouds (\cite{lad74, lad76, wil99, wil03, pov09}): one is located at north side of the \HII \ region (also known as cloud A), and the other is located at south-east side (also known as cloud B).
In this paper, we refer these clouds as N cloud and SW cloud, respectively, and define the boundary of the clouds as shown in Figure \ref{fig:results_peak}b.
The peak intensity map of \COal, which generally corresponds to the kinematic temperature because the line is optically thick and under local thermodynamic equilibrium (LTE) condition in most cases, shows both SW cloud and N cloud have an intensity gradient indicates the clouds are heated by the strong UV radiation from NGC 6618 cluster (\cite{stu88, stu90}).
The highest temperature is up to $\sim$60 K at the interface of the molecular cloud and \HII \ region, and the temperature decreases down to $\sim$20 K in the peripheral regions.
Both of the clouds have clear boundaries at the direction of NGC 6618 cluster, that is mainly due to the stellar wind and/or radiation from the cluster (\cite{thr83, rai87}).
The internal structures of the clouds are observed in \CObl \ and \COcl \ maps because of the small optical depth of the lines.
For the SW cloud, several filaments seems to be connected at the center of the cloud ($l=15.011^{\circ}$, $b=-0.673^{\circ}$), which is also observed by high density tracers (\cite{sne86, rai87, stu90}).
These hub-filament structures are commonly observed in active star forming cloud (e.g., \cite{hil11}).
For the N cloud, a couple of waved filaments are observed in \COb \ map, they are extended in parallel toward the \HII \ region.
The interval and amplitude of the waved filaments are estimated to be $\sim$5 pc and $\sim$2 pc by eye, respectively.
The waved structure implies the influence of the magnetic field \citep{uch91}, however, the velocity resolution of FUGIN data is unsatisfactory on further detailed investigation.
The observations of magnetic field and CO observations with higher velocity resolution are required to improve understanding on the waved filaments.

\begin{figure}[t]
 \begin{center}
  \includegraphics[width=\textwidth]{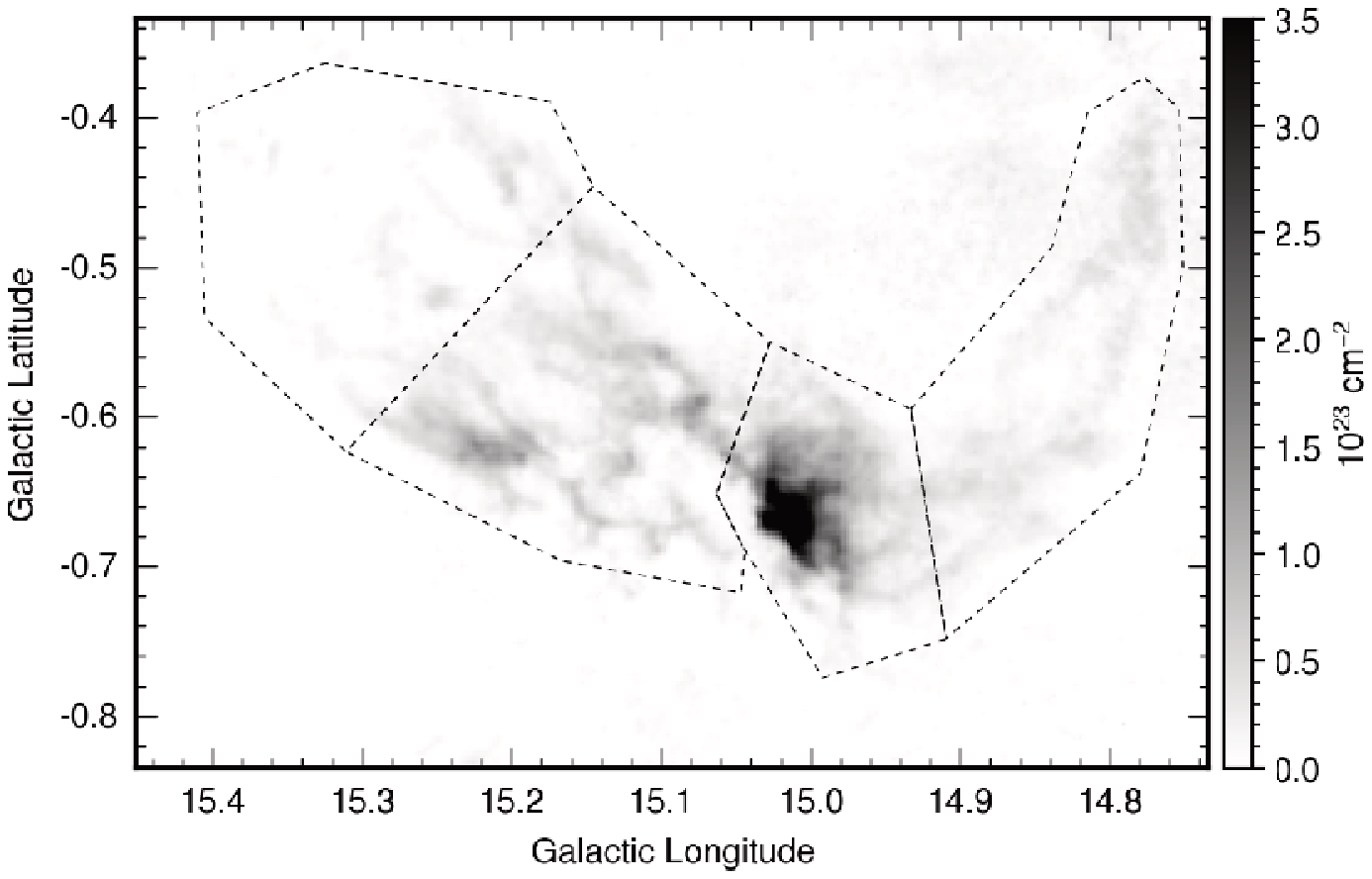} 
 \end{center}
\caption{Column density map of the M17 molecular clouds. Dotted polygons indicate subregions SW1, SW2, N1, and N2.}
\label{fig:results_column}
\end{figure}


\begin{figure}[t]
 \begin{center}
  \includegraphics[width=1\textwidth]{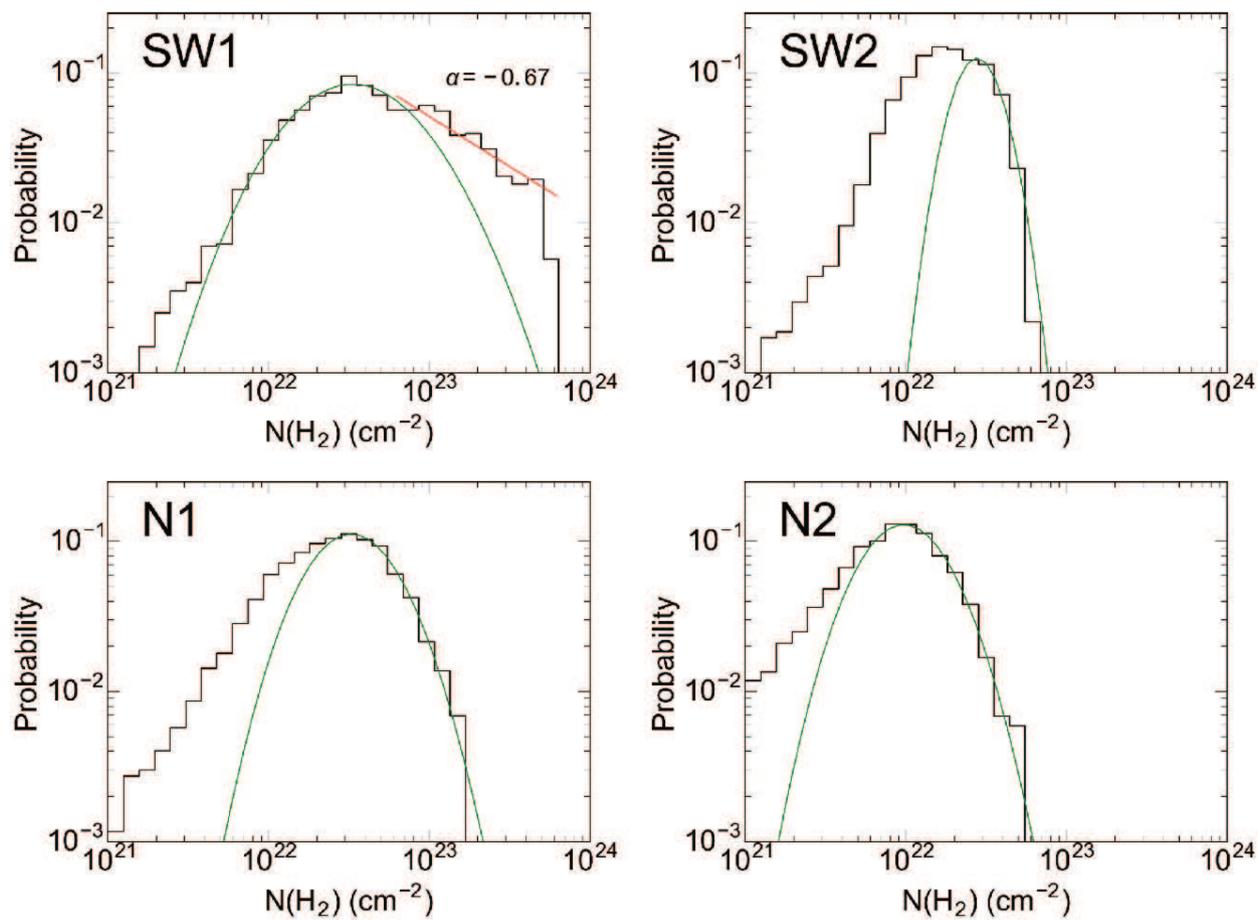} 
 \end{center}
\caption{PDFs for the four subregions in the M17, labbeled as SW1, Sw2, N1 and N2. The green and red lines indicate the results of best fit of log-normal function and power-law function, respectively.}
\label{fig:pdf_sub}
\end{figure}

The column density of the molecular gas is derived by \CObl \ intensity assuming local thermodynamic equilibrium (e.g., \cite{dic78, nis15}).
The excitation temperature, $T_{\rm ex}$, is calculated from \COal \ assuming the line is optically thick.
We adopt the flactional abundance ratio of the \COb \ is $X$[\COb] = 7.1 $\times$ 10$^5$ \citep{fre82}.
Figure \ref{fig:results_column} shows the derived column density map.
The cloud masses are estimated to be $6.5 \times 10^4 M_{\solar}$ and $3.4 \times 10^4 M_{\solar}$ for the SW and N clouds, respectively, by integrating the areas shown in Figure \ref{fig:results_peak}b.
The derived gas masses are roughly consistent with that of \citet{elm76}.
%
Figure \ref{fig:pdf_sub} shows probability density function (PDF) of column density for the four subregions which are defined in the Figure \ref{fig:results_peak}b.
The SW and N clouds are divided into two subregions each, near side and far side to the NGC 6618 cluster, to compare the effect of feedback from the cluster.
All subregions show log-normal distributions at the lower densities with the peak at $\log N \sim 22.0-22.5$.
For both clouds, subregions of near side to the cluster (SW1 and N1) have higher column density than subregions of far side (SW2 and N2).
The SW1 subregion only shows power-law tail at the higher densities with a gradient of $\alpha \sim -0.75$ which derived by least squrare fitting using the data above $\log N = 22.8$.
The PDF has a cut-off at $\log N \sim 23.8$, it is considered mainly because the effect of depletion of the CO molecules \citep{sch16}.

\begin{figure*}[t]
 \begin{center}
  \includegraphics[width=1\textwidth]{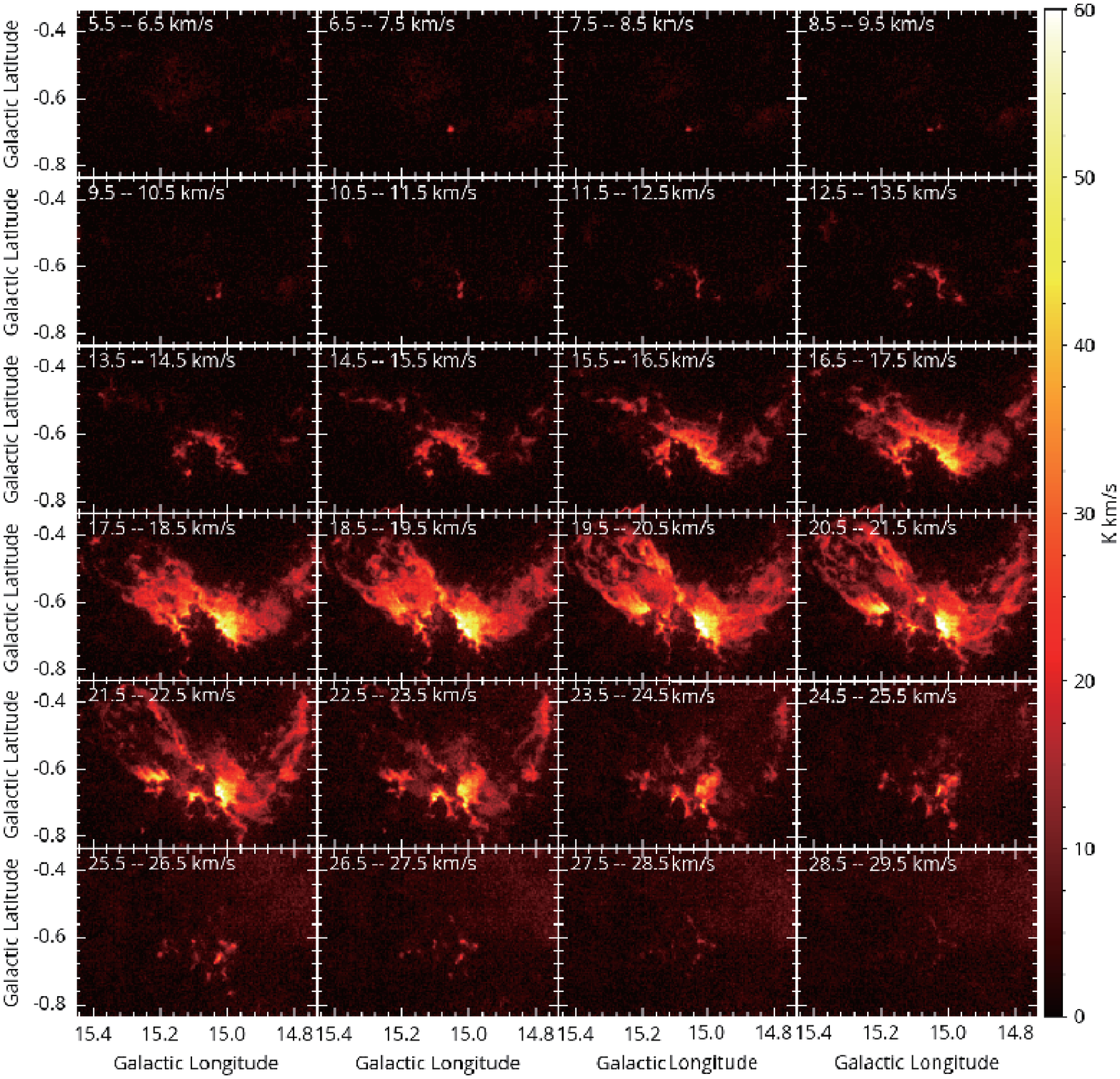} 
 \end{center}
\caption{\COal \ velocity channel maps for the velocity range 5.5 km s$^{-1}$ $<$ \vlsr $<$ 29.5 km s$^{-1}$ made every 1.0 km s$^{-1}$. The center velocity for the integration is indicated in the topleft corner of each panel.}
\label{fig:results_ch12}
\end{figure*}

\begin{figure*}[t]
 \begin{center}
  \includegraphics[width=1\textwidth]{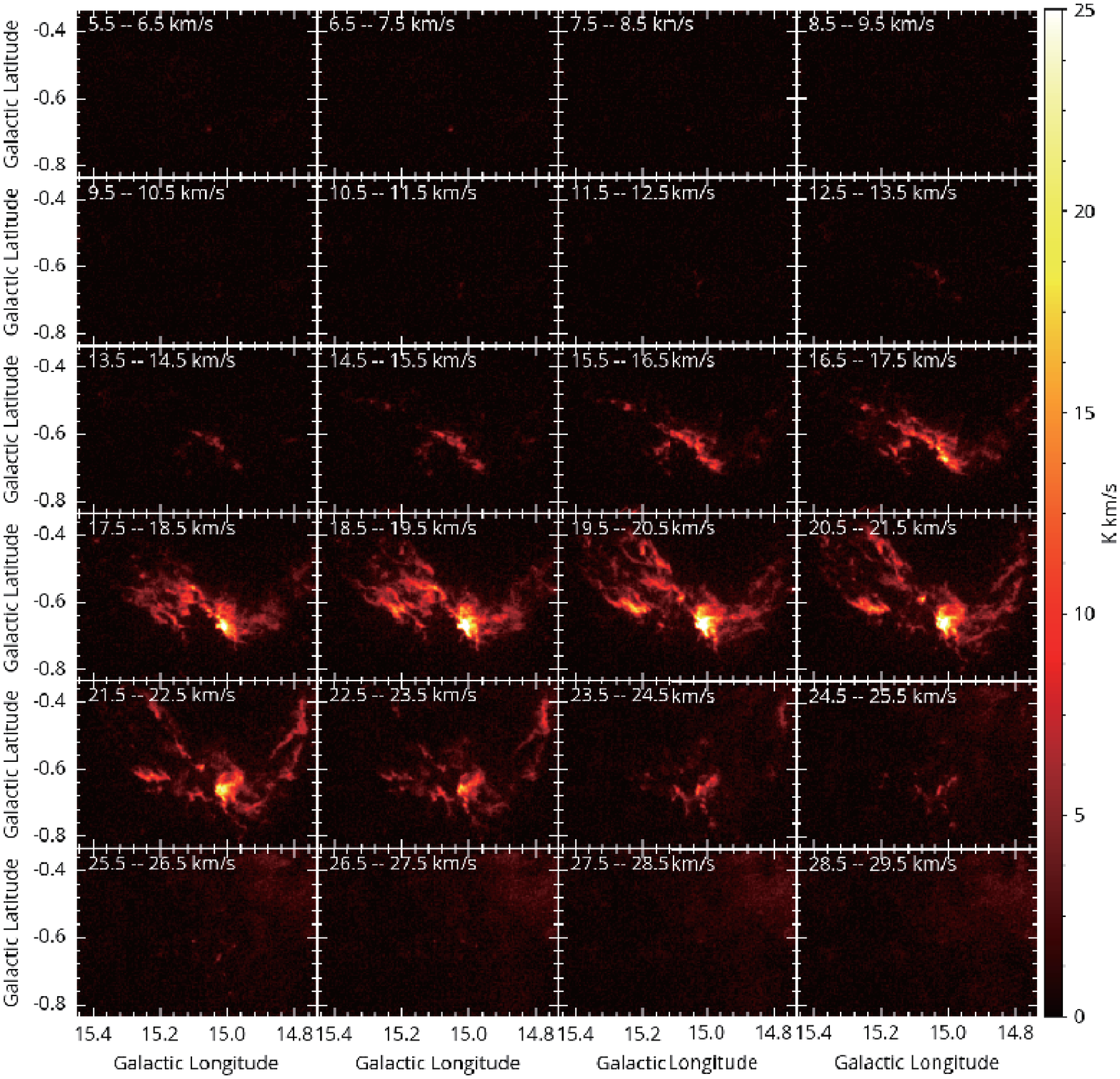} 
 \end{center}
\caption{Same as Figure \ref{fig:results_ch12}, but for \CObl.}
\label{fig:results_ch13}
\end{figure*}

\subsection{Velocity structure}
\label{sec:velocity}

Velocity structures are complicated in the M17 molecular cloud complex, as seen in the velocity channel maps shown in Figures \ref{fig:results_ch12} and \ref{fig:results_ch13}.
The M17 molecular clouds can be divided into four different velocity structures: the very low velocity clump (VLV clump; \vlsr \ = 5.5 -- 9.5 km s$^{-1}$), the low velocity component (LVC; \vlsr \ = 9.5 -- 17.5 km s$^{-1}$), the main velocity component (MVC; \vlsr \ = 17.5 -- 21.5 km s$^{-1}$), and the high velocity component (HVC; \vlsr \ = 21.5 -- 29.5 km s$^{-1}$).
The VLV clump is located in the \HII \ region ($l=15.056^{\circ}$, $b=-0.694^{\circ}$).
The size and mass of the VLV clump are 0.24 pc and 9.6 $M_{\solar}$, respectively.
The VLV clump is spatially connected with lower velocity part of the LVC.
The LVC is located in the surroundings of the \HII \ region.
The boundary of the LVC corresponds to the interface with the \HII \ region, and the cavity in the LVC corresponds to the \HII \ region.
The filamentary structure is elongated with north-south direction from the SW cloud to the N cloud.
The weak \COal \ emission is distributed in the interface with the \HII \ region at the lower velocity channels (\vlsr \ = 9.5 -- 13.5 km s$^{-1}$) of LVC, suggests the gas is a part of the LVC and accelerated by the feedback from the cluster.
The MVC contains most of the mass of the M17 molecular clouds and has a mixed emission distribution of LVC and HVC.
The \COal \ maps show uniformly diffused intensity distribution except for the gradient due to the heating of cluster, whereas the \CObl \ maps show internal structures clearly: several waved filaments in the N cloud and hub-filament structure in the SW cloud.
The HVC is mainly consists of four clumps (see also Figure \ref{fig:results_ch12}): a part of the SW cloud (hereafter HVC-SW), a cloud on the \HII \ region (HVC-SE), eastern part of the N cloud (HVC-NE) and western part of the N cloud (HVC-NW).
The boundary of the HVC-SW corresponds to the interface with the \HII \ region, whereas the HVC-NE and HVC-NW are located away from the interface with the \HII \ region.
The cavities are distributed between the HVC clumps.
The cavity between the HVC-SW and HVC-SE corresponds to the \HII \ region, while that of the HVC-NW and HVC-NE is located at the inner part of the N cloud.
The temperature gradient in the HVC-NE and  HVC-NW indicates feedback from the cluster reaches to inner part of the N cloud through the cavity.
The weak \COal \ emission is distributed in the interface with the \HII \ region at the higher velocity channels (\vlsr \ = 25.5 -- 29.5 km s$^{-1}$) of HVC suggests the gas is a part of the HVC and accelerated by the feedback from the cluster.

\begin{figure*}[t]
 \begin{center}
  \includegraphics[width=1\textwidth]{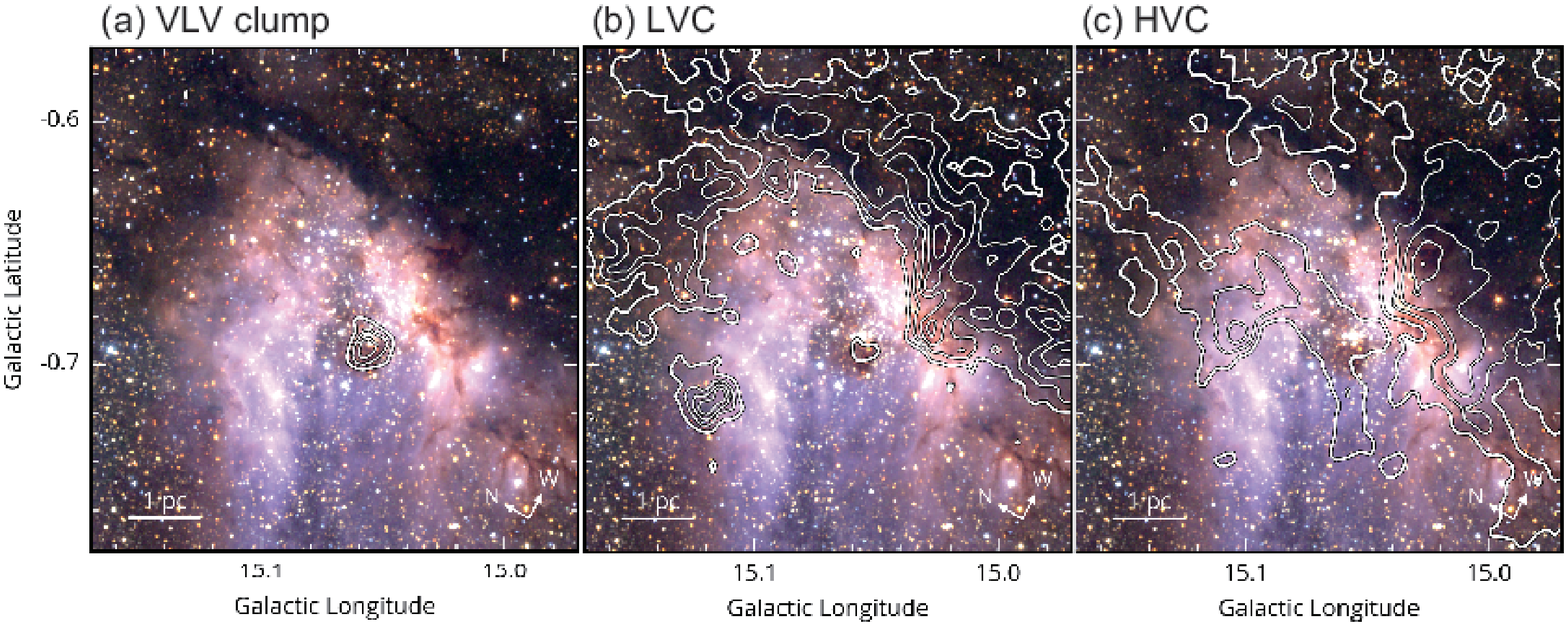} 
 \end{center}
\caption{Close-up images of NGC 6618 cluster from the 2MASS data. The J, H and K$_{s}$ bands are shown in blue, green and red, respectively. Contour indicates \COal \ emission of (a) the VLV clump at 5.5--9.5 km s$^{-1}$ with contour levels for every 23 K km s$^{-1}$, (b) the LVC at 9.5--13.5 km s$^{-1}$ with contour levels for every 27 K km s$^{-1}$, and (c) the HVC at 21.5--29.5 km s$^{-1}$ with contour levels for every 65 K km s$^{-1}$.}
\label{fig:results_2mass}
\end{figure*}

Figure \ref{fig:results_2mass} shows enlarged view of the cluster from 2MASS infrared image overlaid with contours of \COal \ emission of the VLV clump, the LVC, and the HVC.
The VLV clump corresponds to the infrared dark clump that obscures bright diffuse gas on the \HII \ region.
Therefore, the VLV clump is located at the foreground of the \HII \ region.
The LVC is distributed complementary with the infrared bright region, and correlated with infrared dark region that obscures stars behind the GMC.
Therefore, the LVC is located at the same distance as the \HII \ region.
The distribution of HVC shows no correlation with the infrared image, that indicates the HVC is located at far side of the \HII \ region.

\begin{figure*}[t]
 \begin{center}
  \includegraphics[width=1\textwidth]{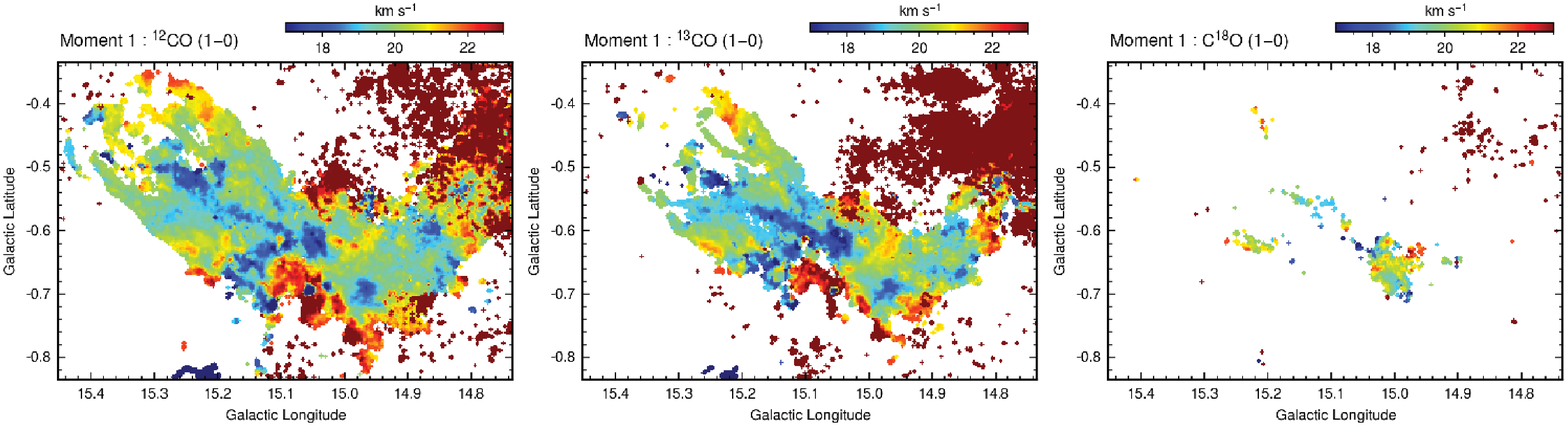} 
 \end{center}
\caption{Intensity-weighted mean velocity map in the velocity range from 16 to 23 km s$^{-1}$ for (left) \COal, (center) \CObl, and (right) \COcl.}
\label{fig:results_mom1}
\end{figure*}

\begin{figure*}[t]
 \begin{center}
  \includegraphics[width=1\textwidth]{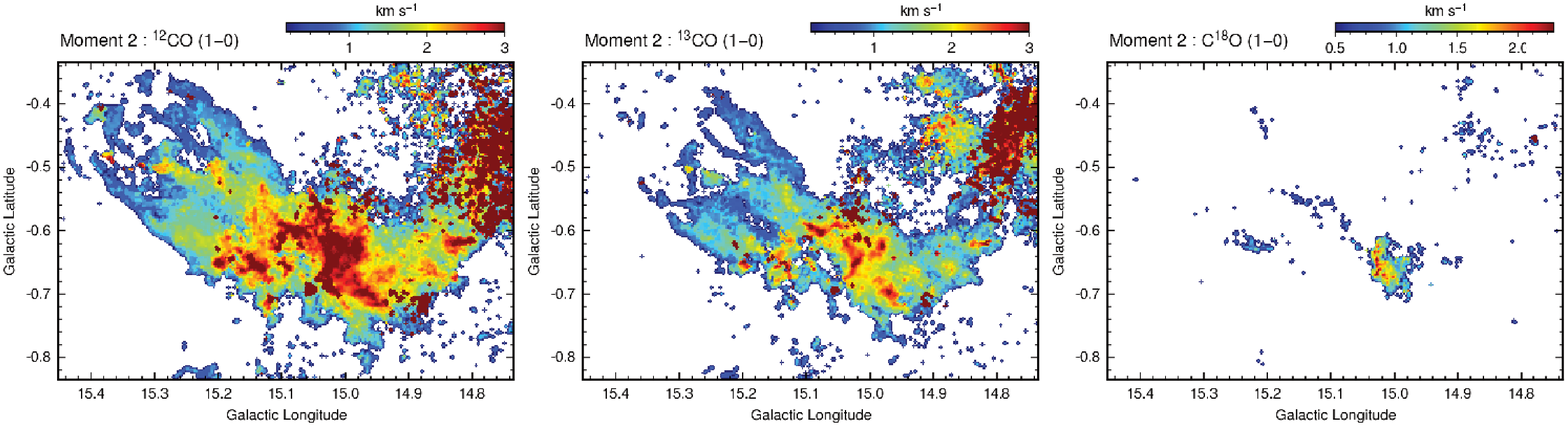} 
 \end{center}
\caption{Velocity dispersion maps of (left) \COal, (center) \CObl, and (right) \COcl.}
\label{fig:results_mom2}
\end{figure*}

Figure \ref{fig:results_mom1} shows the intensity-weighted mean velocity maps of \COal, \CObl, and \COcl.
All maps exhibit quite similar velocity distributions.
Higher velocity gas distributes in outside of the molecular clouds.
The N cloud consists of several filaments that has diferrent velocities.
Figure \ref{fig:results_mom2} shows the velocity dispersion maps.
Larger values are observed around the \HII \ region, however, these are mostly due to exist of the multi velocity components.

\subsection{Feedback from NGC 6618 cluster}
\label{sec:feedback}

\begin{figure}[t]
 \begin{center}
  \includegraphics[width=1\textwidth]{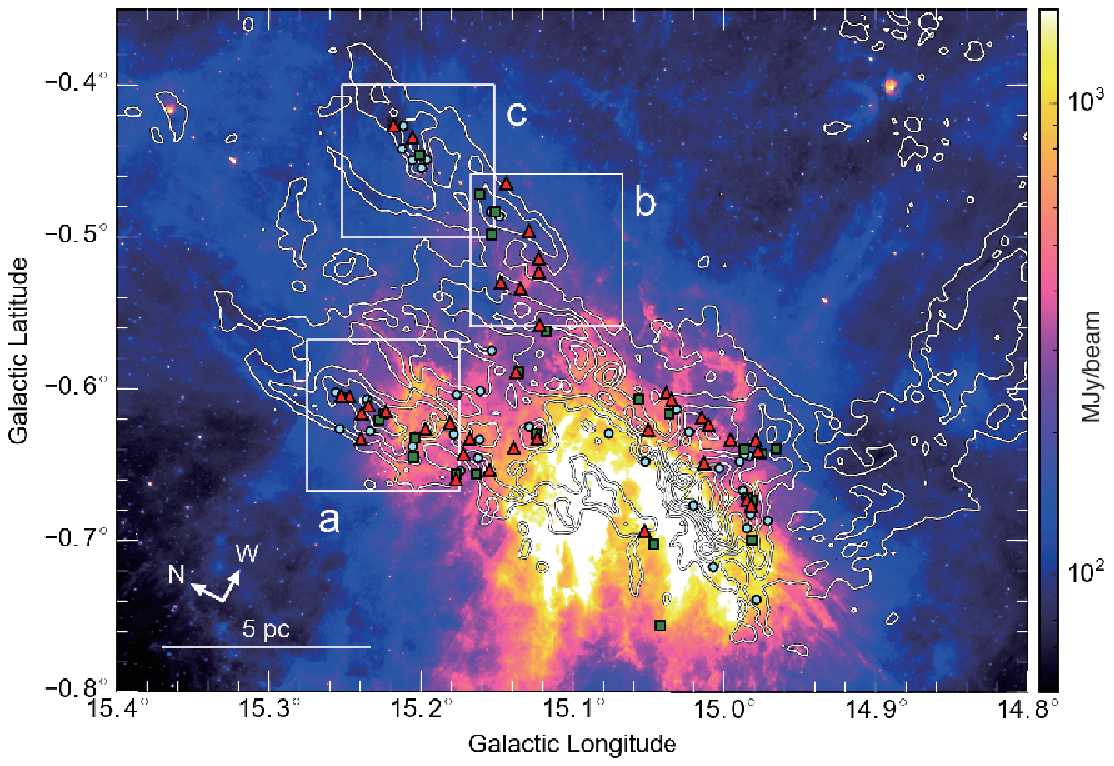} 
 \end{center}
\caption{Spitzer 8 $\rm{\mu m}$ image with a contour of peak temperature map of \CObl. Contours are plotted at every 4 K from 4.5 K, White rectangles indicate subregions where UV interacts with molecular gas in N cloud. Red triangles, green squares and cyan circles show the YSOs class 0/I, II/III and ambiguous, respectively \citep{pov13}.}
\label{fig:8um}
\end{figure}

\begin{figure}[t]
 \begin{center}
  \includegraphics[width=1\textwidth]{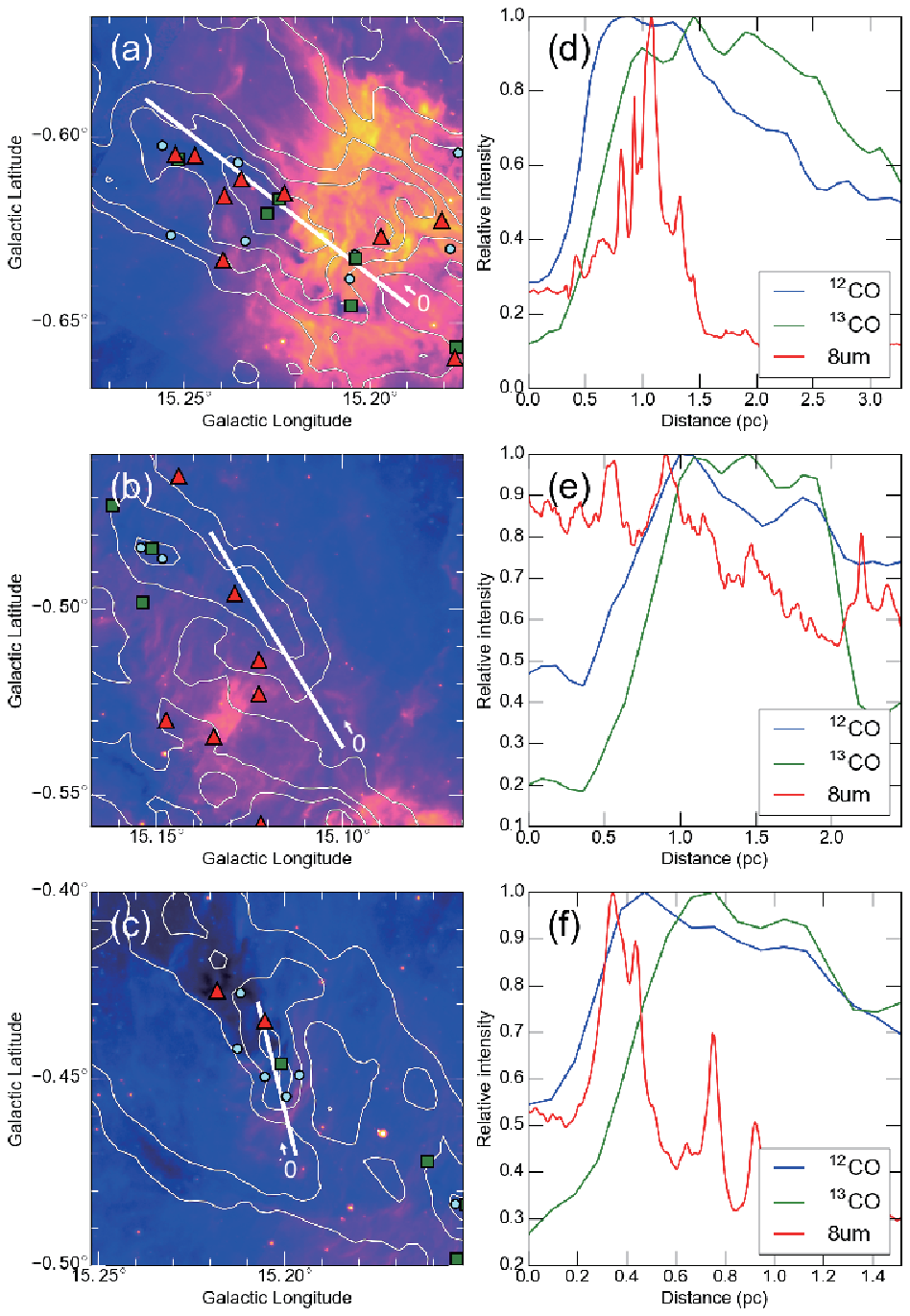} 
 \end{center}
\caption{Left panels: Zoom up views of Figure \ref{fig:8um}. Spitzer 8 $\rm{\mu m}$ image with a contour of peak temperature map of \CObl. Markers indicate same as Figure \ref{fig:8um}. Right Panels: Relative intensities of the \COal, \CObl \ and 8 $\rm{\mu m}$ emissions on the white line which indicated in the left panels.}
\label{fig:N_8um}
\end{figure}

M17 molecular clouds are heavily interacted with the feedback from NGC 6618 cluster.
In this subsection, we describe the relationship of the feedback and molecular gas distributions.

The SW cloud is located adjacent to the cluster and heavily interacting with its feedback at almost all velocity range (Figure \ref{fig:results_ch12}).
Because the interface layer of the \HII \ region and the molecular cloud is generally quite thin, the clear edge shape observed in \COal \ emission indicates that the interface is located as almost edge-on \citep{thr83}.
In \COa \ map, the temperature peak is located at the interface in the SW cloud ($l=15.011^{\circ}$, $b=-0.682^{\circ}$), whereas the peaks are located at the inner region for \COb \ ($l=15.011^{\circ}$, $b=-0.673^{\circ}$) and \COc \ ($l=15.013^{\circ}$, $b=-0.666^{\circ}$).
For all lines, the boundary of the intensity is clear at the interface region, whereas diffuse emission is located behind the peak position, suggests the effect of the feedback is limited to the surface area of the SW clouds.

Figure \ref{fig:8um} shows Spitzer $8 \rm{\mu m}$ intensity image \citep{pov07} which represents distribution of photodissociation region (PDR) where the UV interacts with molecular gas.
The strong $8 \rm{\mu m}$ emission is mainly associated with the interface of \HII \ region and the cloud.
For the SW cloud, the $8 \rm{\mu m}$ intensity is drop down immediately, whereas for N cloud, moderate intensity ($\sim 5 \times 10^{2}$ MJy sr$^{-1}$) emission exists $\sim5$ pc behind the boundary of molecular gas.
The result indicates that the UV feedback from NGC 6618 cluster is limited on the interface region for the SW cloud, whereas the low volume filling factor caused by its velocity structure (see \S \ref{sec:velocity}) allows the UV photons to reach inner part for the N cloud.
We find three $8 \rm{\mu m}$ bright rim structures on filaments in the N cloud that indicated as a, b, and c in Figure \ref{fig:8um}.
Figures \ref{fig:N_8um} a--c show zoom up views of the structures and d--f show relative intensities of CO and $8 \rm{\mu m}$ on the white lines indicated in the panels a--c.
For all of the regions, $8 \rm{\mu m}$ is bright on the edge of the filament toward the direction to NGC 6618 cluster.
Additionally, most of YSOs are located on the filament with bright rim structure in the N cloud (see Figure \ref{fig:8um}).
The number of class 0/I and II/III YSOs in the bright rim filament a--c are (8, 5), (6, 3), and (2, 1), respectively.

%% file: s4_discussion.tex
\section{Discussion}

\subsection{Cavities in the HVC: Possible evidences for cloud-cloud collision}
\label{sec:res-ccc}

\begin{figure*}[t]
 \begin{center}
  \includegraphics[width=\textwidth]{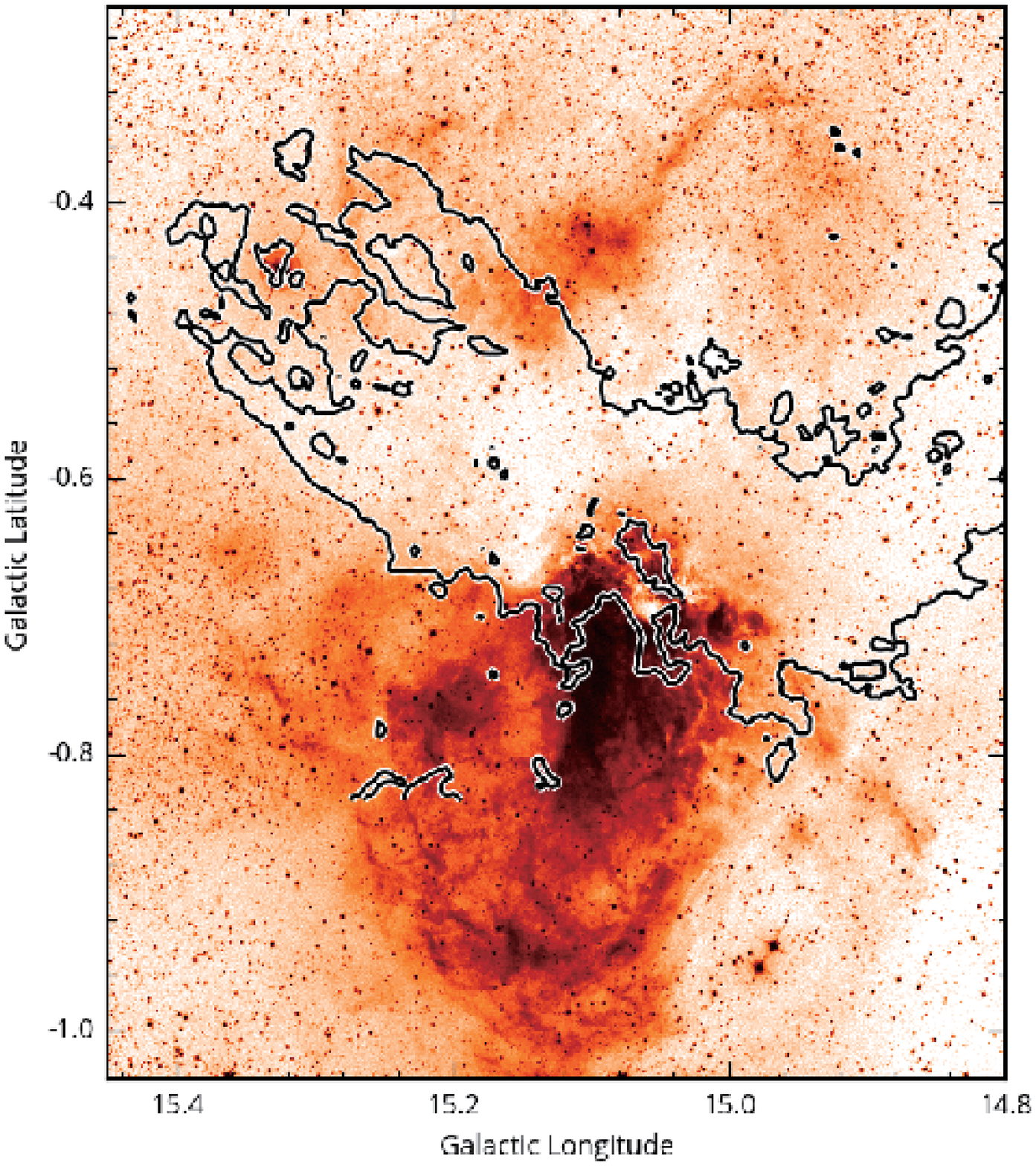}
 \end{center}
\caption{An optical image of the M17 from DSS Red data. Contours indicate the region where the peak temperature of \COal \ is 12 K as a reference of the GMCs.}
\label{fig:dss-red}
\end{figure*}

\begin{figure*}[t]
 \begin{center}
  \includegraphics[width=\textwidth]{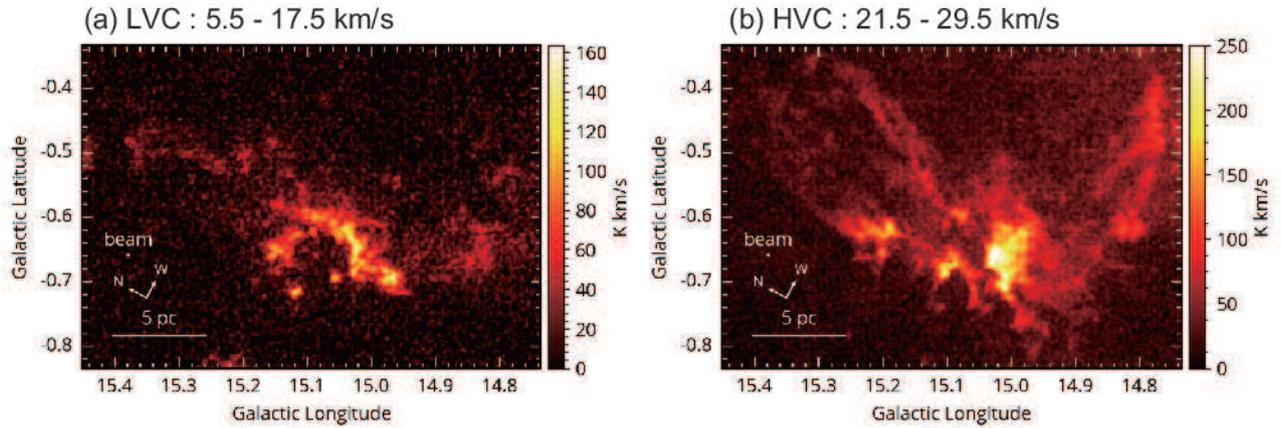}
 \end{center}
\caption{Integrated intensity maps of \COal \ for (a) the LVC (9.5--13.5 km s$^{-1}$) and (b) the HVC (21.5--29.5 km s$^{-1}$).}
\label{fig:lvc-hvc}
\end{figure*}

\begin{figure*}[t]
 \begin{center}
  \includegraphics[width=\textwidth]{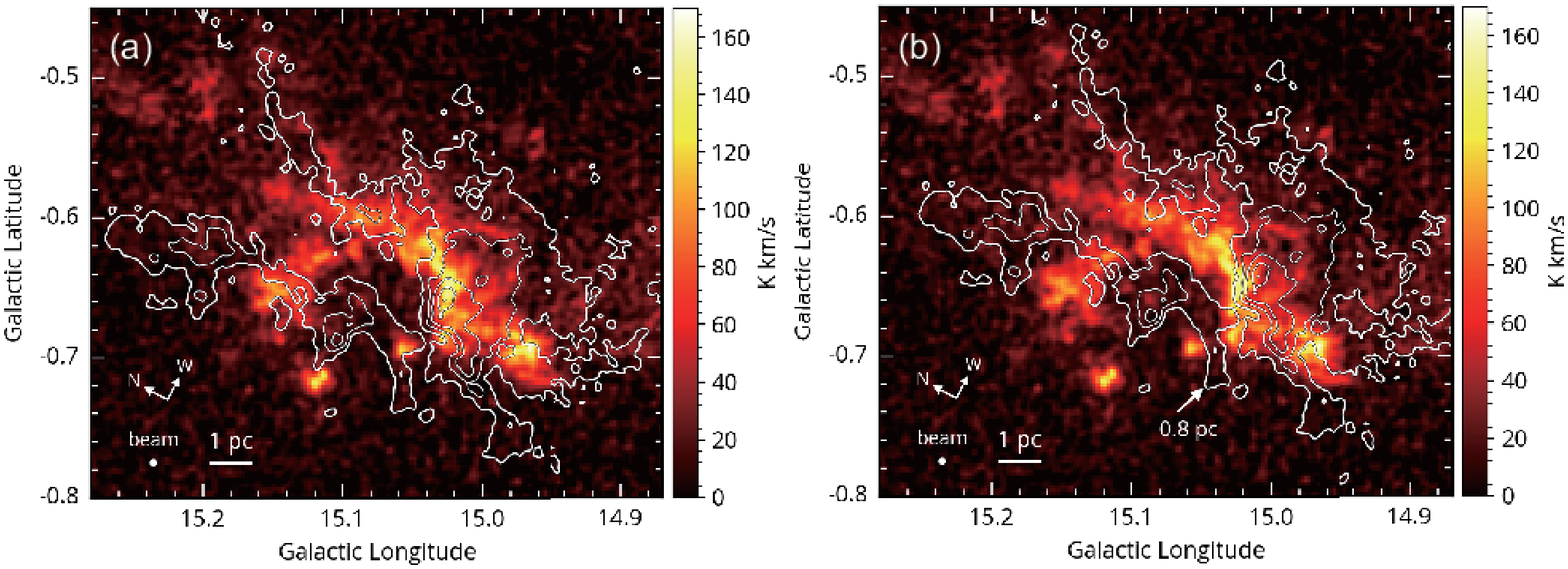}
 \end{center}
\caption{(a) The complementary distributions of the two velocity distributions. Image indicates \COal \ emission of the LVC at 9.5--13.5 km s$^{-1}$ and contour indicates the HVC at 21.5--29.5 km s$^{-1}$ with contour levels for every 65 K km s$^{-1}$. (b) Same as (a), but contours are displaced 0.8 pc as shown the arrow.}
\label{fig:ccc}
\end{figure*}

The cavity in the LVC shows clear correspondence with the \HII \ region, whereas that of HVC shows little.
In the N cloud, UV photons through the HVC cavities and interact with the molecular gas which is located at inner part of the cloud (\S \ref{sec:feedback}).
Ionized gas should be observed at the north edge of the cavity if the cavity is opened by the cluster feedback.
An image from DSS Red band, which has a sensitivity for emission of H$\alpha$ line, shows most ionized gas are located at the southeast of the GMCs, while no significant emission are distributed at the north of the N cloud (Figure \ref{fig:dss-red}).
Therefore, the feedback of the cluster is unlikely to form the HVC cavities.

Here, we propose the possibility that the cavities were created by collisions of the molecular clouds.
Cloud collisions can make cavities by density difference of the colliding cloud pair.
The higher density cloud penetrates the lower density cloud, and creates the cavity in the lower density cloud.
In such a case, the higher- and lower-density clouds show complementary distributions each other, which can be found in different velocity channels if colliding cloud pair have different line-of-sight velocities.
The complementary distributions in different velocity channels are observed in star forming GMCs (e.g., \cite{fur09, dob14}), and in numerical simulations (e.g., \cite{hab92, tak14, mat15}).
In the case of M17, the width of the HVC cavities are similar to the width of LVC filaments (Figure \ref{fig:lvc-hvc}), while the HVC cavities are mismatched to the LVC filaments spatially (Figure \ref{fig:ccc}a).
The cloud collision makes the spatially matched complementary distribution if clouds collided on the direction of line-of-sight.
However, if the collision direction is off line-of-sight, the collided clouds move away each other on the projected plane \citep{fuk17}.
Therefore, the spatial mismatch of the HVC cavities and the LVC filaments can be explained as the mismatch of the collision direction and line-of-sight.
The displacement of the HVC toward the LVC is estimated to be 0.8 pc in east-southeast direction by eye (Figure \ref{fig:ccc}b).
The LVC is located as surrounding the east side of the HVC for the SW cloud, while the LVC and the HVC are distributed in complementary for the \HII \ region and N cloud.
All cavities in the LVC have counterparts in the HVC emission, and vice versa.
Hence, collisions of the LVC and the HVC can explain the formation of the cavities.
In this case, the cavities are firstly created by the cloud collisions, and then the feedback of the cluster interacts with the remnant of the collision that further expands the cavities.
The result that the LVC is located at the same distance with the \HII \ region and the HVC is located far side of the \HII \ region (see  \S \ref{sec:velocity}), is consistent with this collision scenario.

The traveling distance of the HVC cloud and collision velocity are estimated to be 1.1 pc and 9.9 km s $^{-1}$, respectively, if we assume the collision direction toward line-of-sight is 45$^{\circ}$.
In this case, the collision timescale, which is derived from the traveling distance divided by the collision velocity, is estimated to be $1.1 \times 10^{5}$ yr.
The collision timescale is equivalent to that of NGC 6618 cluster (\cite{hof08}; see detail in \S \ref{sec:cluster-formation}).
This timescale is comparable to that of RCW 38, which is the youngest super star cluster in the Galaxy, and its time scale is estimated to be $\sim 1 \times 10^5$ yr \citep{fuk16}.
Because of its youth and small impact area ($\sim 0.5$ pc) of the collision, RCW 38 cluster is heavily obscured by natal molecular cloud, even though the cloud harbors a number of  intense ionization sources.
By contrast, M17 has much larger cavity than that of RCW 38.
This difference can be explained naturally in the collision scenario: the cavities of the LVC and HVC are the results of the larger impact area ($\sim 5$ pc) of the collision than that of RCW 38.

\subsection{Triggered formation of NGC 6618 cluster by cloud-cloud collision}
\label{sec:cluster-formation}

The cloud-cloud collision is possible in M17 GMCs as discussed in \S \ref{sec:res-ccc}.
In this subsection, we discuss the possibility of triggered formation of NGC 6618 cluster by the collision.

The \HII \ region is heavily ionized by NGC 6618 cluster which contains $>53$ OB stars \citep{hof08} including O4-O4 binary system CEN1 \citep{chi80, rod12}.
If we assume the O stars are formed by a constant mass accretion rate on the collision timescale $1.1 \times 10^{5}$ years, the mass accretion rate is required to be $1.1 \times 10^{-3} M_{\solar}$ yr$^{-1}$ for $120 M_{\solar}$ star (two O4 stars).
MHD numerical simulations of the cloud-cloud collision have shown that the turbulence is enhanced and the magnetic field is amplified in the interface layer, and the mass accretion rate is enhanced by two orders of magnitude to $10^{-4}$--$10^{-3} M_{\solar}$ yr$^{-1}$ as compared with the pre-collision state \citep{ino13}.
The collision velocity of 9.9 km s$^{-1}$ in the M17 clouds is comparable to that in the simulations, therefore, this collision event possibly provides such a high mass accretion rate.
\citet{fuk16} pointed out that super star clusters are formed in the colliding clouds which has column density larger than $10^{23}$ cm$^{-2}$, and single O star is formed in lower column density case.
The column density of M17 SW reaches $4 \times 10^{23}$ cm$^{-2}$ at the peak position, therefore, M17 has a potential to form a super star cluster by the cloud collision, if the hypothesis of the column density threshold is correct.

The stellar populations of NGC 6618 is clearly divided into two groups from the color-magnitude diagram \citep{hof08}.
One group is on a zero-age-main-sequence curve and the other is on a side of younger than $10^{5}$ yr isochrone curve.
This generation gap indicates that an trigger event occurred in the star forming GMC to form younger stellar member $10^{5}$ yr ago.
Column density PDF for the SW cloud (Figure \ref{fig:pdf_sub}) shows a power-law tail, implies that spontaneous star formations are ongoing in entire the SW cloud.
Furthermore, the timescale of younger member of NGC 6618 ($\sim 10^{5}$ yr) is in good agreement with the timescale of the collision ( $\sim 10^{5}$ yr).
Hence, we conclude that the scenario of cloud-cloud collision likely well explains the stellar population and its formation history of NGC 6618 cluster.

%% file: s5_conclusions.tex
\section{Conclusions}

We have observed CO $J=1-0$ lines toward the entire extent of the M 17 molecular cloud with the highest angular resolution as a part of FUGIN.
The main conclusions of the present study are summarized as follows.

\begin{enumerate}

\item
The M17 molecular clouds can be divided into four different velocity structures: the very low velocity clump (VLV clump; \vlsr \ = 5.5 -- 9.5 km s$^{-1}$), the low velocity component (LVC; \vlsr \ = 9.5 -- 17.5 km s$^{-1}$), the main velocity component (MVC; \vlsr \ = 17.5 -- 21.5 km s$^{-1}$), and the high velocity component (HVC; \vlsr \ = 21.5 -- 29.5 km s$^{-1}$).
These three components are associated with M17 \HII \ region.
The VLV clump, LVC, and HVC are located at foreground, same distance, and far side of the \HII \ region, respectively.

\item
The feedback of NGC 6618 cluster interacts with the LVC and the HVC.
The accelerated diffuse gas are detected in lower velocity part of the LVC and higher velocity part of the HVC at the interface of \HII \ region.

\item
The LVC and the HVC have cavities.
UV photons from NGC 6618 cluster through the cavities and interact with the molecular gas, which is located at inner part of the cloud.
This interaction is well correlated with YSOs in N cloud.

\item
The cavities are well explained as a results of cloud collision of the LVC and the HVC.
The LVC and the HVC are distributed in complementary after the HVC is displaced by 0.8 pc in east-southeast direction.
The collision velocity and timescale is estimated to be 9.9 \kms \ and $1.1 \times 10^{5}$ yr, respectively.

\item
The scenario of cloud-cloud collision likely well explains the stellar population and its formation history of NGC 6618 cluster proposed by \citet{hof08}.
The high collision velocity (9.9 \kms) can provide the high mass accretion rate up to 10$^{-3}$ $M_{\solar}$ yr$^{-1}$.
The high column density ($4 \times 10^{23}$ cm$^{-2}$) of the SW cloud might arrow to form a massive cluster such as NGC 6618.
The timescale of younger member of NGC 6618 ($\sim 10^{5}$ yr) is in good agreement with the timescale of the collision ( $\sim 10^{5}$ yr).

\end{enumerate}

\begin{ack}
 
This work was supported by JSPS KAKENHI grant numbers,
15K17607
.
The authors would like to thank the members of the 45-m group of Nobeyama Radio Observatory for support during the observation. 
The Nobeyama 45-m radio telescope is operated by Nobeyama Radio Observatory, a branch of National Astronomical Observatory of Japan.
Data analysis was carried out on the open use data analysis computer system at the Astronomy Data Center, ADC, of the National Astronomical Observatory of Japan.
This research made use of astropy, a community-developed core Python package for Astronomy \citep{ast13}, in addition to NumPy and SciPy \citep{wal11}, Matplotlib \citep{hun07} and IPython\citep{per07}.

\end{ack}